\documentclass[runningheads]{llncs}
\usepackage[T1]{fontenc}
\usepackage{amsmath,amssymb,amsfonts}%
\usepackage{graphicx}%
\usepackage{multirow}%

\usepackage{mathrsfs}%
\usepackage[title]{appendix}%
\usepackage{xcolor}%
\usepackage{textcomp}%
\usepackage{manyfoot}%
\usepackage{booktabs}%
\usepackage{algorithm}%
\usepackage{algorithmicx}%
\usepackage{algpseudocode}%
\usepackage{listings}%
\usepackage{makecell}
\usepackage{bbm}
\usepackage{lipsum}
\newcommand\blfootnote[1]{%
  \begingroup
  \renewcommand\thefootnote{}\footnote{#1}%
  \addtocounter{footnote}{-1}%
  \endgroup
}
\begin{document}
\title{Geo-UNet: A Geometrically Constrained Neural Framework for Clinical-Grade Lumen Segmentation in Intravascular Ultrasound}
%
%
\author{Yiming Chen\inst{1}\textsuperscript{*} 
\and Niharika S. D'Souza \inst{2}\textsuperscript{*}  \and Akshith Mandepally \inst{3} \and Patrick Henninger\inst{3} \and Satyananda Kashyap\inst{2} \and Neerav Karani \inst{1} \and Neel Dey \inst{1} \and Marcos Zachary \inst{3} \and Raed Rizq \inst{3} \and Paul Chouinard \inst{3} \and Polina Golland \inst{1} \and Tanveer F. Syeda-Mahmood \inst{2}}
%

\authorrunning{Chen Y., D'Souza N.S. et al.} 
%
%

\institute{Massachusetts Institute of Technology, Boston, MA, USA
\and
IBM Research, Almaden, San Jose, CA, USA
\and
Boston Scientific Corporation, Maple Grove, MN, USA
}
%
%
%
\maketitle              
\begin{abstract}
Precisely estimating lumen boundaries in intravascular ultrasound (IVUS) is needed for sizing interventional stents to treat deep vein thrombosis (DVT). Unfortunately, current segmentation networks like the UNet lack the precision needed for clinical adoption in IVUS workflows. This arises due to the difficulty of automatically learning accurate lumen contour from limited training data while accounting for the radial geometry of IVUS imaging. We propose the Geo-UNet framework to address these issues via a design informed by the geometry of the lumen contour segmentation task. We first convert the input data and segmentation targets from Cartesian to polar coordinates. Starting from a convUNet feature extractor, we propose a two-task setup, one for conventional pixel-wise labeling and the other for \textit{single boundary} lumen-contour localization. We directly combine the two predictions by passing the predicted lumen contour through a new activation (named CDFeLU) to filter out spurious pixel-wise predictions. Our unified loss function carefully balances area-based, distance-based, and contour-based penalties to provide near clinical-grade generalization in unseen patient data. We also introduce a lightweight, inference-time technique to enhance segmentation smoothness. The efficacy of our framework on a venous IVUS dataset is shown against state-of-the-art models. \blfootnote{\textsuperscript{*} indicates joint first-authorship}
\end{abstract}

\section{Introduction}
Deep Vein Thrombosis (DVT) is a serious condition that can cause significant short-term discomfort and lead to irreversible venous system damage that may be limb or life-threatening~\cite{scarvelis2006diagnosis}. It is a precursor to pulmonary embolism, a critical condition where a clot travels to the lungs, impeding blood oxygenation. To manage DVT, clinicians often utilize Intravascular Ultrasound (IVUS)~\cite{secemsky2022intravascular} to guide endovascular treatments, where a catheter equipped with an ultrasound transducer is inserted to visualize internal structures and pinpoint anatomical landmarks. IVUS samples are organized into pullbacks, where consecutive frames of images are captured as the catheter travels through the blood vessel, emitting sound waves that are reflected by/pass through structures based on their densities~\cite{secemsky2022intravascular}. The physician may remove the thrombus and insert balloons or stents in its place to keep the vessel open. These devices are sized based on nearby healthy regions, where accurate measurement of the vessel's lumen is crucial for avoiding complications like pain from improper device sizes or fatal stent migration~\cite{stahr1996importance}. Automatic segmentation of venous IVUS (v-IVUS) images is challenging owing to variability/irregularity in tissue/vessel appearance across subjects due to thin vessel walls, noise, stents, artifacts, and the manual nature of the pullback (i.e. variable longitudinal frame rate across pullbacks due to manual control of the catheter by the physician).

Deep Neural Networks (DNNs) for vascular segmentation~\cite{arora2023state} have soared in popularity due to their ability to provide improved performance without manual intervention during deployment. 
Variants of the UNet~\cite{ronneberger2015u} have been successful for plaque/calcification detection and vessel segmentation~\cite{arora2023state,xie2020two,huang2023post,blanco2022fully} as well as stent~\cite{wissel2022cascaded} and lesion detection/classification~\cite{meng2023deep} for coronary artery disease. These use either 2D images~\cite{blanco2022fully} or 3D image blocks~\cite{huang2023post,kashyap2023feature} as inputs and produce a pixel-wise map of the segmentation target as the output. The IVUS segmentation literature focuses on arterial acquisitions which provide a different field of view (FoV) and use a motorized pullback providing a fixed longitudinal frame rate. However, venous acquisitions are not well-studied, and most existing techniques do not generalize well to v-IVUS data due to under/over-segmentation of lumen regions in the presence of imaging artifacts and their predilection to output spurious, fragmented predictions when there are nearby vessels or tissue structures. We posit that this is due to their inability to reflect the radial geometry of the imaging modality and constrain the output to be a single contiguous lumen region, as dictated by the anatomy under consideration. 

In this paper, we alleviate the issues above by designing a new neural framework, named Geo-UNet---a fully convolutional architecture for lumen segmentation from venous IVUS images that satisfies radial contour-geometry constraints directly through the imaging-representation, architecture, and loss functions (as opposed to imposing anatomical constraints via regularization~\cite{oktay2018ACNN} or architecture~\cite{citation-0,d2021matrix,d2023mspd,nandakumar2020multi} alone). Our method features 3 main components: \textbf{1) Input representation:} we operate on 2D-image inputs converted from Cartesian to polar coordinates which better reflect inherent IVUS imaging physics~\cite{szarski2021improved,blanco2022fully}. \textbf{2) Anatomically Constrained Self-informing Network}: We propose a two-task setup with a shared UNet feature extraction module. In polar space, the lumen boundary is a single contour. While the natural prediction target is a standard pixel-level segmentation, we design a second objective to predict a single lumen boundary contour. Using this prediction as a guide, we refine the pixel-level segmentation via a new activation function---\textit{CDFeLU}, based on the cumulative distribution function. This regularization mitigates spurious predictions from pixel-level segmentation without the need for additional post-processing, a known shortcoming of prior approaches. During training, our unified loss function combines area-based, distance-based, and contour-based penalties to provide improved generalization. \textbf{3) Inference-time Continuity Enhancement}: Based on the radial geometry in imaging and properties of the convolutional UNet, we propose a continuity enhancement technique, coined Geo-UNet++, which is a lightweight, inference-time procedure to address wrap-around discontinuities at $0/2\pi$ angles in the segmentation estimation. Our framework compares favorably against state-of-the-art segmentation baselines with consistent improvements in segmentation Dice scores and derived lumen diameter estimation for stent sizing. 

\section{Geo-UNet for Lumen Segmentation from v-IVUS}
\begin{figure}[t!]
\centering
\fbox{\includegraphics[scale=0.22]{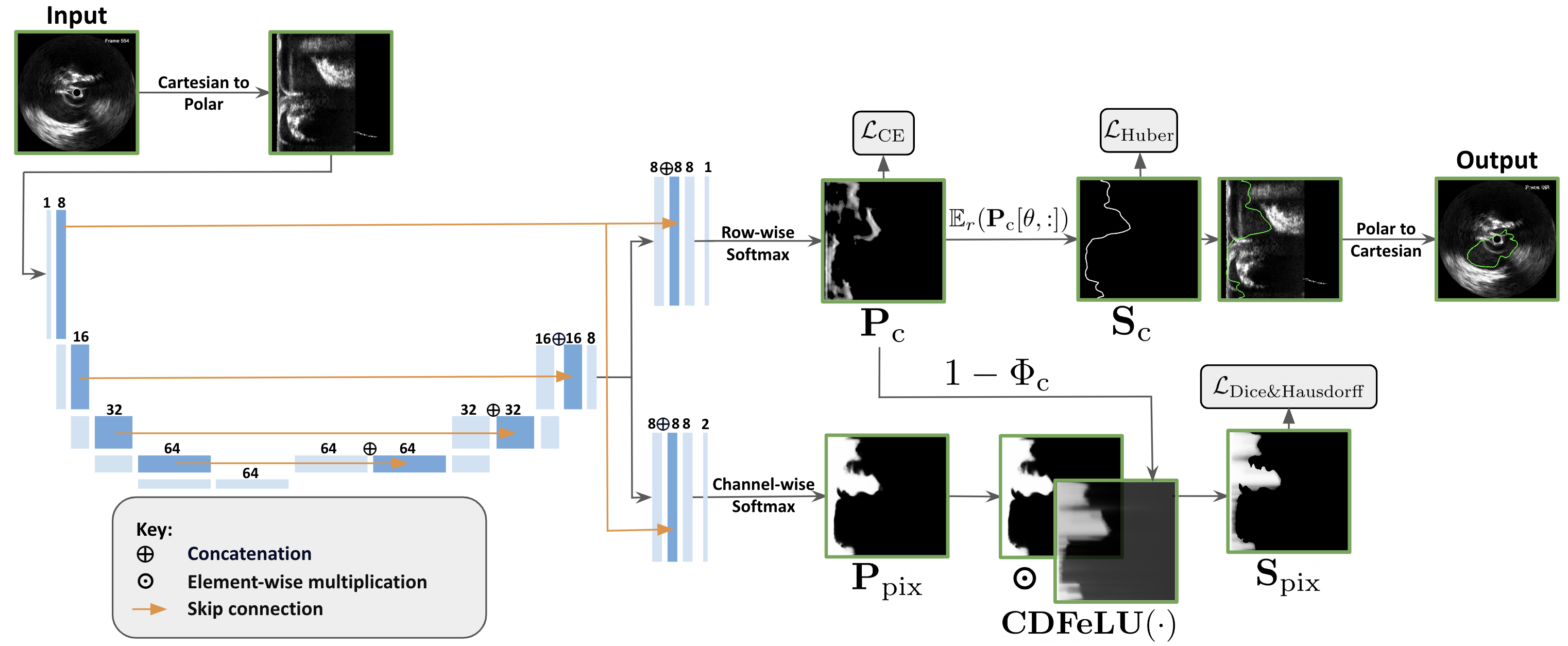}}
\caption{\footnotesize \textbf{Geo-UNet Architecture for Lumen Segmentation}: The feature extractor is a fully convolutional UNet module with inputs of polar 2D IVUS frames. The top branch produces a probability map for the lumen contour ($\mathbf{P_{\text{c}}}$) via a row-wise softmax, which is converted to a single contour segmentation ($\mathbf{S_{\text{c}}}$) via a row-wise expectation function. The bottom branch produces a per-pixel probability map ($\mathbf{P_{\text{pix}}}$) via a channel-wise softmax. $\text{CDFeLU}(\cdot)$ allows the top branch to inform the bottom, refining the pixel-wise probabilities to give the segmentation ($\mathbf{S_{\text{pix}}}$) that is compared against the (polar) ground-truth lumen mask. The loss functions are highlighted in grey.} 

\label{GeoUNet}
\end{figure}
Fig.~\ref{GeoUNet} illustrates the Geo-UNet framework. The shared convUNet feature extractor connects to two prediction branches as detailed below.

\medskip
\noindent\textbf{Lumen Contour Estimation Branch:} 
In polar space, the horizontal and vertical axes of an IVUS image correspond to radii ($r$) and angles ($\theta$), respectively. $\mathbf{Y}_{\text{pix}}$ denotes the ground-truth binary lumen mask of size $R\times R$ (R=256 pixels). Summing along the $r$ coordinate for each $\theta$ gives the contour lumen map $\mathbf{Y}_{c}[\cdot]$ of size $R\times 1$. $\mathbf{Y}_{c}[\theta] = \sum_{r}{\mathbf{Y}_{\text{pix}}[\theta,r]}$ captures the lumen depth at each $\theta$, a distinct value in \{0,\dots, R\}. The lumen boundary is a single, smooth contour with no self-intersection (i.e. has a distinct depth $r \in \{0,\dots, R-1\}$ for each $\theta \in \{0,\dots, R-1\}$, after discretizing the range $[0,2\pi]$ into R intervals). The top network branch captures the lumen contour by computing a softmax across each row of the single-channel output to obtain a row-parse probability map $\mathbf{P}_\text{c}$ of size $R\times R$. The entries $\mathbf{P}_\text{c}[\theta,r] \in [0,1]$  denote the probability that the contour depth at $\theta$ is $r$ and is ideally high along the lumen contour and near $0$ elsewhere. We convert $\mathbf{P}_\text{c}$ into a segmentation contour $\mathbf{S}_{\text{c}}$ with an expectation across radii values: $\mathbf{S}_{\text{c}}[\theta] = \mathbbm{E}_{r}(\mathbf{P}_\text{c}[\theta,:]) = \sum_{r = 0}^{R-1} r*\mathbf{P}_\text{c}[\theta,r]$, accounting for the uncertainty along boundary pixels in a differentiable operation~\cite{He2019}. We use two training losses for $\mathbf{P}_\text{c}$ and $\mathbf{S}_\text{c}$. First, we compute the cross entropy between $\mathbf{P}_\text{c}$ and $\mathbf{Y}_{c}$:
\begin{equation}
\mathcal{L}_\text{CE}=\frac{-1}{R^2}\sum_{\theta,r = 0,0}^{R-1}\mathbbm{1}[\mathbf{Y}_{c}[\theta] = r]\textbf{log}(\mathbf{P}_\text{c}[\theta,r]) + \mathbbm{1}[\mathbf{Y}_{c}[\theta] \neq r]\textbf{log}(1-\mathbf{P}_\text{c}[\theta,r])
\label{sp1}
\end{equation}
We then encourage $\mathbf{S}_{\text{c}}[\theta]$ to match $\mathbf{Y}_{c}[\theta]$ using a Huber loss~\cite{huber1964robust}:
\begin{eqnarray}
\mathcal{L}_\text{Huber}(\cdot) = \sum_{\theta=0}^{R-1} \frac{d_\theta^2}{2} \mathbbm{1}(\vert{\mathbf{d}_\theta}\vert< 1) + (\vert{\mathbf{d}_\theta}\vert-0.5)\mathbbm{1}(\vert{\mathbf{d}_\theta}\vert\geq 1),
\label{sp2}
\end{eqnarray}
where $\mathbf{d}_{\theta} = \mathbf{Y}_{c}[\theta] - \mathbf{S}_\text{c}[\theta]$. To get a polar binary segmentation that guarantees a single lumen in Cartesian space, for each $\theta$, we have 1s for pixels to the left of/along $\mathbf{S}_\text{c}[\theta]$ and 0 elsewhere. This serves as the final prediction output.

\medskip
\noindent\textbf{Pixel-wise Segmentation Branch with Probabilistic Contour Maps:} Applying a conventional channel-wise softmax operation~\cite{arora2023state}, the bottom branch outputs a pixel-wise probability map $\mathbf{P}_\text{pix}$ of size $R\times R$, where $\mathbf{P}_\text{pix}[\theta,r]$ denotes the probability that pixel $[\theta,r]$ is in or on the lumen boundary. To reconcile this with the lumen contour estimate, we compute a dense probability map from $\mathbf{P}_{\text{c}}$ via a novel activation function based on the cumulative distribution function (CDF). Let $\Phi_\text{c}[\theta,r] = \text{CDF}(\mathbf{P}_\text{c}[\theta,r])$, the transformation $(1-\Phi_\text{c}[\theta,r])$ models the confidence that the pixel $[\theta,r]$ is contained within the lumen that is larger at smaller radii, serving as a probabilistic mask for $\mathbf{P}_\text{pix}$. We compute the refined pixel-wise segmentation $\mathbf{S}_\text{pix}$ of size $R\times R$ via the activation $\text{CDFeLU}(\mathbf{P}_\text{pix}, \mathbf{P}_\text{c})$:
\begin{equation*}
\mathbf{S}_\text{pix}[\theta,r] = {\mathbf{P}_\text{pix}[\theta,r]}*{(1-\Phi_\text{c}[\theta,r])
 = \mathbf{P}_\text{pix}[\theta,r]*\Big[1 - \sum_{j=0}^r \mathbf{P}_\text{c}[\theta,j]\Big]}
\end{equation*}
CDF error Linear Units (CDFeLU) is analogous to Gaussian Error Linear Units (GELU)~\cite{hendrycks2016gaussian}, where the CDF error is estimated based on the geometry of the lumen boundary as opposed to a normal distribution. Finally, we impose a combination of area-based (Dice) and distance-based (Hausdorff~\cite{Cardoso2022}) losses on $\mathbf{S}_\text{pix}$ to match the ground-truth pixel-wise lumen mask $\mathbf{Y}_{\text{pix}}$:
\begin{equation}
\mathcal{L}_{\text{Dice\&Hausdorff}}(\cdot) = \lambda * \mathcal{L}_{\text{Dice}}(\mathbf{S}_\text{pix}, \mathbf{Y}_{\text{pix}}) + (1-\lambda)* \mathcal{L}_{\text{Haus.}}(\mathbf{S}_\text{pix}, \mathbf{Y}_{\text{pix}})
\label{dense}
\end{equation}
with the trade-off $\lambda \in (0,1)$ determined experimentally to be 0.9. Note that by design, $\text{CDFeLU}(\mathbf{P}_{\text{pix}}, \mathbf{P}_\text{c})$ de-emphasises regions outside the lumen (right of $\mathbf{\mathbf{Y}_{c}[\theta]}$), filtering out potentially spurious predictions in $\mathbf{P}_{\text{pix}}$, a task usually reserved for manual/semi-automated post-processing. At the same time, it reinforces overlaps between $\mathbf{P}_{\text{c}}$ and $\mathbf{P}_{\text{pix}}$, effectively encouraging Geo-UNet to focus on estimates that align well across the two branches during training. 

\begin{figure}[h]
\fbox{\includegraphics[scale=0.35]{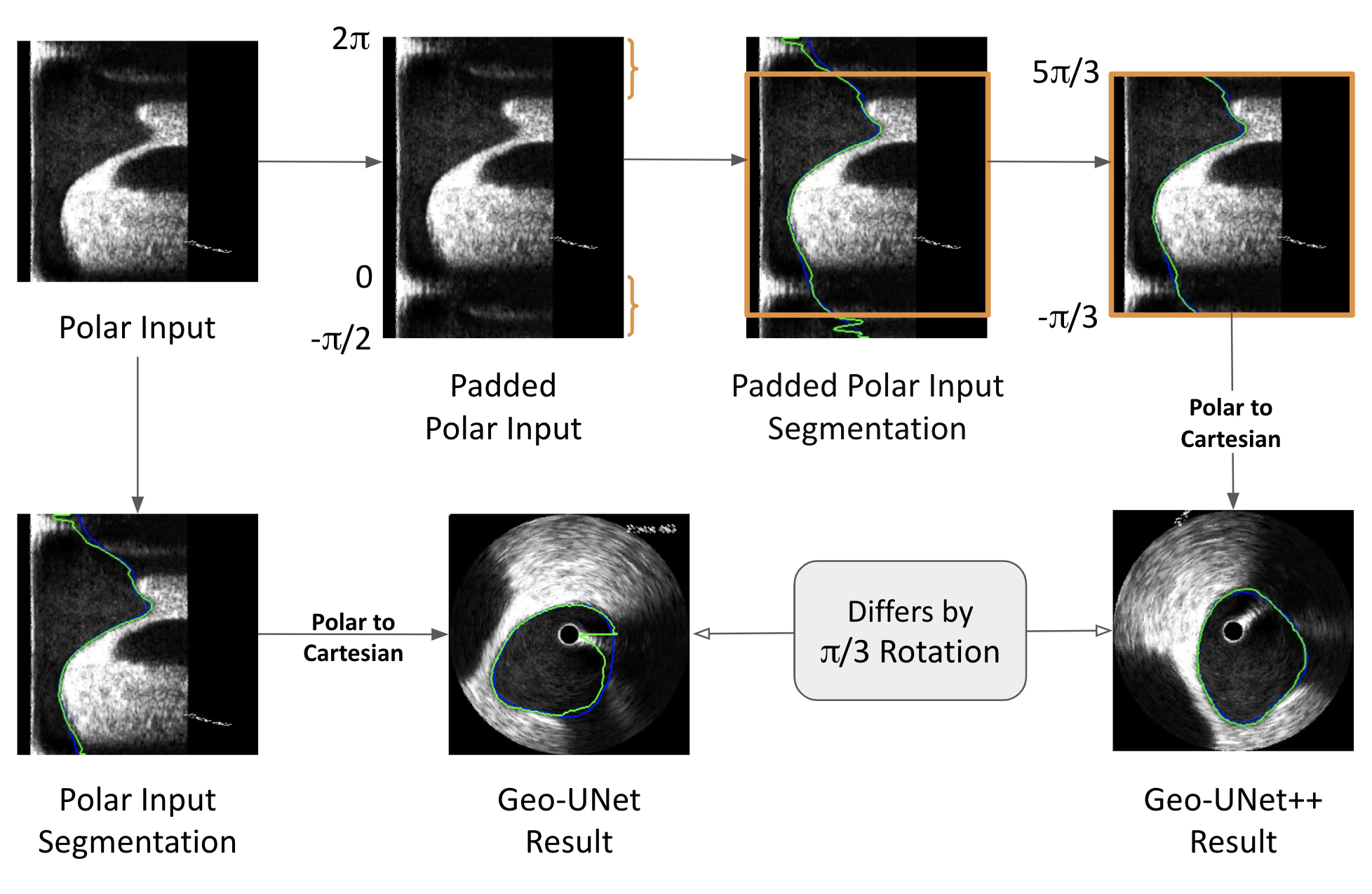}}
\caption{\textbf{Geo-UNet++: Inference-time Segmentation Enhancement:} The bottom middle image shows the performance of Geo-UNet when given a polar input image. The green is the prediction, and the blue is the ground truth. In the bottom left image, note the sudden jumps and misalignment in the green prediction at the top and the bottom of the image, corresponding to $0$ and $2\pi$, respectively, yielding discontinuity at $0$/$2\pi$ in the Cartesian representation. Starting from the top left input image and to the right, we illustrate the ideas behind Geo-UNet++. Exploiting convolution's lack of dependence on input dimensions, we perform inference using the \textit{same} trained Geo-UNet model, on an input wrap-padded with a repetition of the top of the original input, as highlighted by the orange braces. The padding provides additional context near 0. To recover the segmentation, we take the middle portion from $-\frac{\pi}{3}$ to $\frac{5\pi}{3}$ which typically avoids border discontinuities at the top and bottom of the padded image. This is essentially segmentation prediction on the Cartesian input image rotated counterclockwise by $\frac{\pi}{3}$ and does not affect the clinical objective of diameter estimation from the segmentation mask. On the lower right, we see that the Geo-UNet++ result is smoother and nearly perfectly aligned with the ground truth.} 
\label{GeoUNet++}
\end{figure}

\medskip
\noindent\textbf{Geo-UNet++ to Alleviate Wrap-around Artifacts during Inference:} Recall that we map pixel intensities from Cartesian space to r-$\theta$ space to generate polar images, where $\theta \in \{0,\dots, 2\pi\}$. A consequence is that the intensities of the model predictions are not constrained to align at $\theta=0$ and $\theta=2\pi$, as they lie at the top and bottom borders of the polar image. This often results in a wrap-around discontinuity when converting back to Cartesian coordinates that consistently induces errors in the diameter estimation. To alleviate this, we introduce an inference-time technique based on the radial nature of the Cartesian v-IVUS images and properties of convolution. We apply vertical wrap-padding to yield a rectangular, continuous input ranging $\theta=\{-\pi/2,\dots,2\pi\}$ by copying over the additional $\pi/2$ context. With frozen weights, the trained Geo-UNet model can be applied as is to the padded input. We slice the output across the middle section $\theta=\{-\pi/3,\dots,-\pi/3+2\pi\}$) to avoid edge effects in the padded input predictions, before finally presenting the result on the rotated Cartesian input. We observe improved prediction alignment along the re-sliced output for the padded input. See Fig.~\ref{GeoUNet++} (supplementary) for a walk-through. This increases deployment time marginally (0.3-0.4ms/frame) to enhance accuracy. 

\section{Data, Experimental Evaluation, and Results}\label{Results}

\begin{figure}[ht]
\fbox{\includegraphics[scale=0.255]{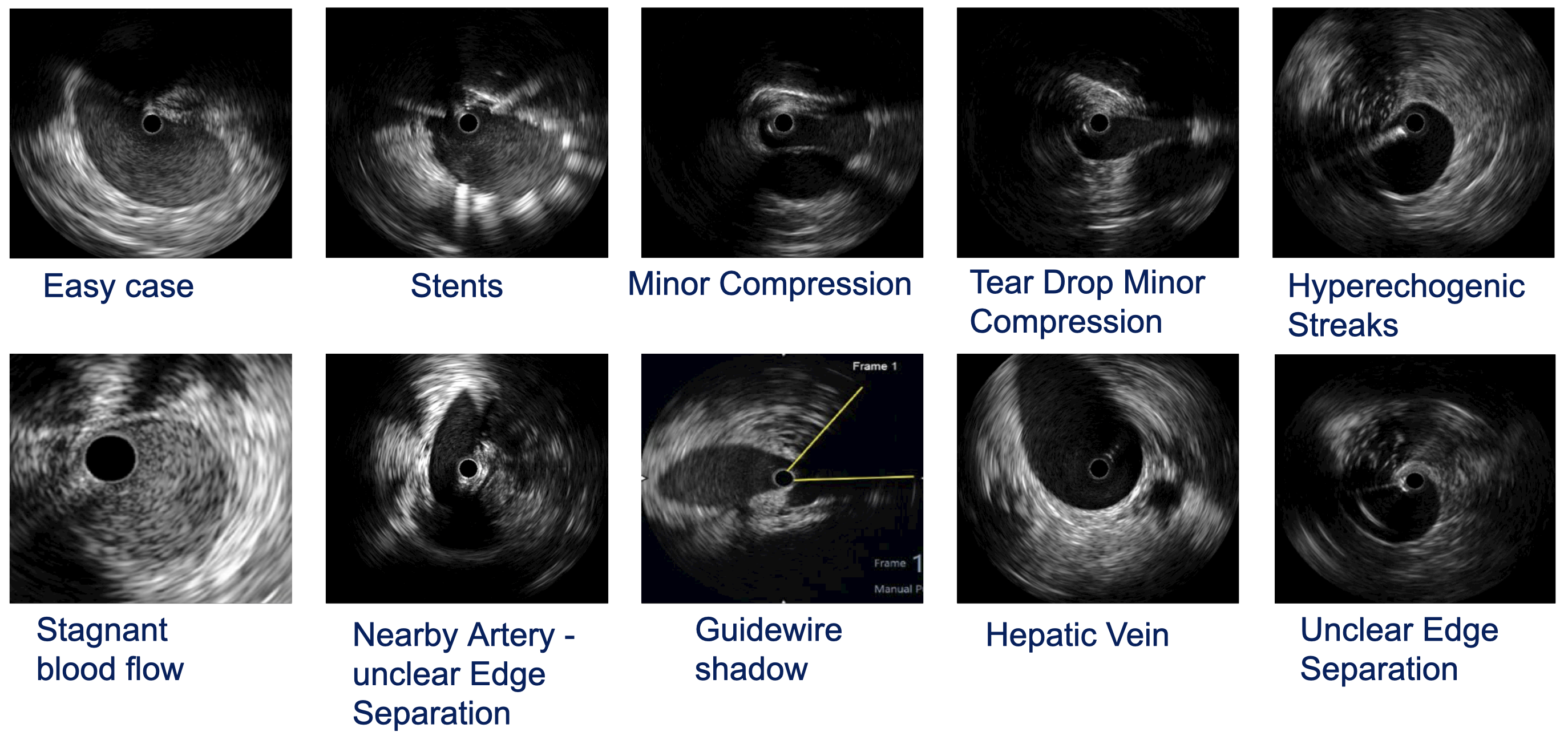}}
\caption{Variations in appearance that are all considered as N1 frames with normal anatomy.} 
\label{N1 variability}
\end{figure}

\medskip
\noindent{\textbf{Data:}} Our images are acquired using the Boston Scientific OC35 peripheral imaging catheter, which uses a rotating transducer to generate cross-sectional views. The catheter has a 70mm imaging diameter and a 15MHz operating frequency. It is typically used in the detection and treatment of venous disease (e.g. DVT, non-thrombotic iliac venous lesions, chronic post-thrombotic syndrome, and more). No registration is needed to align the IVUS frames by the nature of the acquisition. We obtained data for 79 patients with 166 pullbacks of varying durations. The data is labeled per frame and partitioned into two groups: diseased and normal. The former refers to regions with acute/subacute clots and chronic Post Thrombotic Syndrome (PTS). The latter contains labels N1 (frames with typical geometry despite variability in appearance shown in Fig.~\ref{N1 variability} (supplementary)) and N2 (frames with irregular geometry due to compression from nearby vessels but no thrombus present). Since stent-sizing is performed on healthy frames, all N1/N2 frames were labeled by expert annotators, for a total of 77,917 annotated image frames. Given the increased variability in appearance and subjectivity in annotation, the lumen in N2 frames is qualitatively harder to segment as compared to N1 frames.

\medskip
\noindent \textbf{Implementation Details:} We train all models on healthy images (frames marked N1 and N2) and adopt a three-fold cross-validation which stratifies pullbacks across patients (53/21/5 train/test/validation). Input IVUS frames and model outputs are of size $256\times256$ ($R=256$). Hyperparameters across all experiments are determined using the validation set. We use a batch size of 3 with 16 gradient accumulation steps. The Adam optimizer is used with a scheduler that linearly decreases the learning rate from $10^{-4}$ to $10^{-7}$ over 50,000 training iterations. We apply stacked augmentations including rotation, translation, shear, contrast enhancement, Gaussian blur, intensity scaling, and speckle noise on Cartesian inputs for better generalization~\cite{zhang2020augment}. The training loss sums Eqns.(\ref{sp1}-\ref{dense}). To save on compute time, we only retain $\mathcal{L}_{\text{Huber}}(\cdot)$ (Eq.(\ref{sp2})) at each validation step to guide model optimization for Geo-UNet/ablations and $\mathcal{L}_{\text{Dice}}(\cdot)$ for the vanilla U-Net baselines. Our machine has 50 CPU cores and 2 A-100 NVIDIA GPUs with 32GB RAM, resulting in an average training time of 3.5-4 hrs per cross-validation fold. To estimate the lumen diameter from a segmentation mask, we pass lines through the center of mass (COM) of the largest component at 5$^{\circ}$ increments. The longest and shortest lengths of intersection with the mask border are the major and minor diameters, respectively.

Finally, to further encourage spatial contiguity in the predicted masks, we tried introducing implicit smoothness constraints via 1D average pooling on the polar representation across the $\theta$ axis and as a separate post-processing mechanism~\cite{blanco2022fully} on the output. However, both strategies provided negligible performance improvements when weighed against the additional training/inference times.

\medskip
\noindent{\textbf{Evaluation/Clinical Targets:}} In addition to the test-Dice, we evaluate the measurement error in the diameter of the major/minor axes of the predicted lumen against that of the ground-truth lumen~\cite{stahr1996importance}. Commercial stents are sized on N1 frames, are available in 0.5mm increments, and are sized against the average of the major and minor diameter~\cite{stahr1996importance}. Per a clinician, the models need to achieve a major and minor axis diameter error within 0.25/0.5/0.75mm for 50/90/95\% of all N1 frames. N2 frames are mainly used for vessel compression detection and not for stent-sizing. Thus, they have less stringent clinical targets of 50/70\% of frames within errors of 0.5/0.75mm.

\subsection{Baselines Comparisons and Ablations}
We curate our baselines to reflect the state-of-the-art in the fields of medical image segmentation and automated processing of IVUS images.

\medskip
\noindent{\textbf{MedSAM:}} Medical Segment Anything Model~\cite{Ma_2024} is a general-purpose, promptable 2D-segmentation model with a ViT backbone~\cite{xiao2023transformers}, trained on multiple modalities (CT, MRI, ultrasound, etc). The inputs are 2D medical images and a user-specified bounding box to produce a binary pixel-wise segmentation without fine-tuning. We input the Cartesian v-IVUS images and a fixed bounding box based on the FoV to accommodate lumen regions with the largest diameters.

\medskip
\noindent{\textbf{BoundaryReg:}} BoundaryReg is a recent approach~\cite{He2019} based on convolutional UNets that was designed to produce layer surface segmentation for retinal Optical Computed Tomography (OCT). Like GeoUNet, this model estimates both dense pixel and sparse contour predictions using a shared UNet followed by two output convolutional layers without distinct skip connections. It also has additional topology modules to separate retinal layers. As A-scan OCT images and retinal layer segmentation have analogous geometric properties to polar v-IVUS representations and lumen boundary estimation, respectively, we implement this baseline for our application according to the details in~\cite{He2019}.

\medskip
\noindent{\textbf{Cartesian Dice \& Hausdorff:}} Convolutional UNets are commonly used for lumen segmentation from 2D (arterial) IVUS images~\cite{arora2023state}. To adopt these baselines to v-IVUS, we use the architecture from Fig.~\ref{GeoUNet} with only the bottom branch where inputs are Cartesian v-IVUS images and outputs are Cartesian masks. We train using $\mathcal{L}_{\text{Dice\&Hausdorff}}(\cdot)$ between predictions and ground truths~\cite{ReductionHausdorff2019,Cardoso2022}.

\medskip
\noindent{\textbf{Polar Dice \& Hausdorff:}} In line with prior work~\cite{blanco2022fully,szarski2021improved}, we adopt a similar architecture and loss function as the previous baseline, but convert the inputs and targets to polar representations. This baseline also serves as an ablation for Geo-UNet where the contribution of the contour estimation branch is omitted. We obtain a single lumen region from the potentially fragmented pixel-wise predictions by post-processing the outputs to retain the largest connected component~\cite{arora2023state}, both in this approach and the previous baseline.

\medskip
\noindent{\textbf{Ablation Studies:}} To evaluate Geo-UNet, we perform two ablations that systematically remove its key constituent components. These comparisons are (1) Geo-UNet excluding the CDFeLU re-weighting and (2) Geo-UNet without the pixel-wise prediction branch. The former uses the same loss function as Geo-UNet while the latter trains the model on a combination of $\mathcal{L}_{\text{CE}}(\cdot)$ and $\mathcal{L}_{\text{Huber}}(\cdot)$.

\begin{table}[ht!]
\footnotesize{
\caption{Model Performance. Best performance is in bold, second best is underlined.}\label{tab1}
\begin{tabular}{|c|c|c|c|}
\hline
Methodology &  \makecell{Test Dice \\(avg/std)} & \makecell{\% Frames w. Maj. Dia. err.\\
within 0.25/0.50/0.75mm} & \makecell{\% Frames w. Min. Dia. err.\\
within 0.25/0.50/0.75mm}\\
\hline
\multicolumn{4}{|c|}{\textbf{Against Baselines (N1 frames)}} \\
\hline
\textbf{Geo-UNet++}  & \underline{0.95/0.045} & \underline{66}/\textbf{84}/\textbf{90} & \textbf{73}/\textbf{89}/\textbf{94} \\ 
\textbf{Geo-UNet}  & \textbf{0.95/0.034} & \textbf{69}/\textbf{84}/\textbf{90} & 69/85/\underline{91} \\ 
MedSAM~\cite{Ma_2024}  & 0.31/0.087 & 0/0/0  & 0/0/0 \\
BoundaryReg~\cite{He2019}   & 0.94/0.043 & 60/78/86  & \underline{70}/\underline{86}/\underline{91}  \\
Cart. Dice \& Haus.  & 0.93/0.051 & 61/77/83  & 62/79/87  \\
Polar Dice \& Haus.  & 0.94/0.038 & \underline{66}/\underline{80}/\underline{87}  & 67/84/90  \\
\hline
\multicolumn{4}{|c|}{\textbf{Against Baselines (N2 frames)}} \\
\hline
\textbf{Geo-UNet++}  & \textbf{0.88/0.094} & \underline{41}/\underline{59}/\underline{69} & \textbf{60}/\textbf{80}/\textbf{87} \\
\textbf{Geo-UNet}  & \underline{0.87/0.10} & \textbf{47}/\textbf{64}/\textbf{73} & \underline{57}/\underline{76}/\underline{85} \\
MedSAM~\cite{Ma_2024} & 0.23/0.085 & 0/0/0  & 0/0/0  \\
BoundaryReg~\cite{He2019}   & \underline{0.87/0.093} & 36/54/65  & 55/74/84  \\
Cart. Dice \& Haus.  & 0.83/0.12 & 32/44/52  & 44/63/74 \\
Polar Dice \& Haus.   & 0.86/0.12 & 40/58/\underline{69}  & 55/74/83  \\
\hline
\multicolumn{4}{|c|}{\textbf{Against Ablations (N1 frames)}} \\
\hline
\textbf{Geo-UNet}  & \textbf{0.95/0.034} & \textbf{69}/\textbf{84}/\textbf{90} & \textbf{69}/\textbf{85}/\textbf{91} \\
w/o CDFeLU   & 0.94/0.035 & \textbf{69}/\underline{82}/\underline{88}  & \underline{65}/\underline{83}/\underline{90}  \\
w/o pixel-wise pred.  & \underline{0.95/0.039} & 67/81/87  & \textbf{69}/\textbf{85}/\textbf{91} \\
\hline
\multicolumn{4}{|c|}{\textbf{Against Ablations (N2 frames)}} \\
\hline
\textbf{Geo-UNet}  & \underline{0.87/0.10} & \textbf{47}/\textbf{64}/\textbf{73} & \textbf{57}/\textbf{76}/\textbf{85} \\
w/o CDFeLU   & 0.86/0.10 & 45/\underline{63}/\underline{72}  & \underline{53}/\underline{71}/\underline{81}  \\
w/o pixel-wise pred.  & \textbf{0.88/0.092} & \underline{46}/62/71  & \textbf{57}/\textbf{76}/\textbf{85} \\
\hline
\end{tabular}}
\end{table}
\begin{figure}[b!]
\fbox{\includegraphics[scale=0.20]{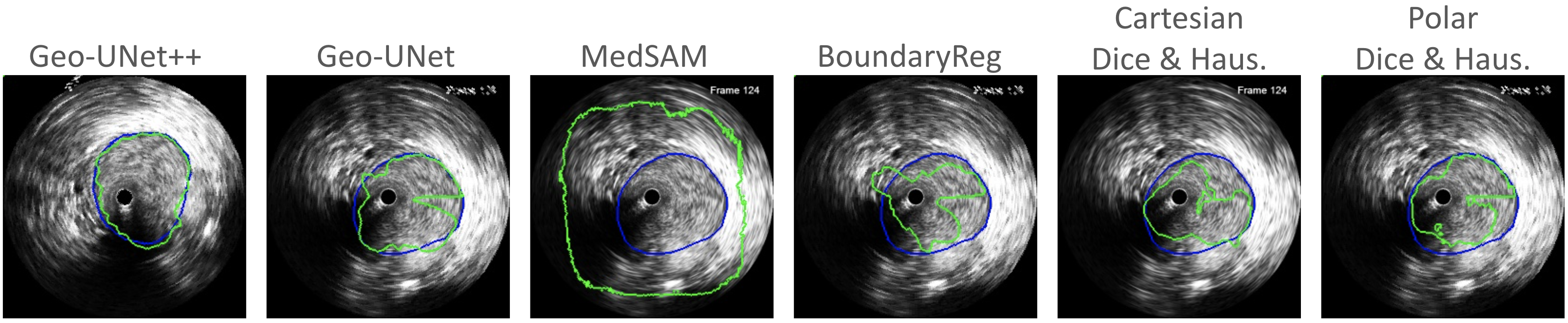}}
\caption{Example of lumen segmentation performance. (Green-Predicted, Blue-Truth)} 
\label{comparison}
\end{figure}

\subsection{Lumen Segmentation Performance Analysis}
\begin{figure}[ht]
\fbox{\includegraphics[scale=0.28]{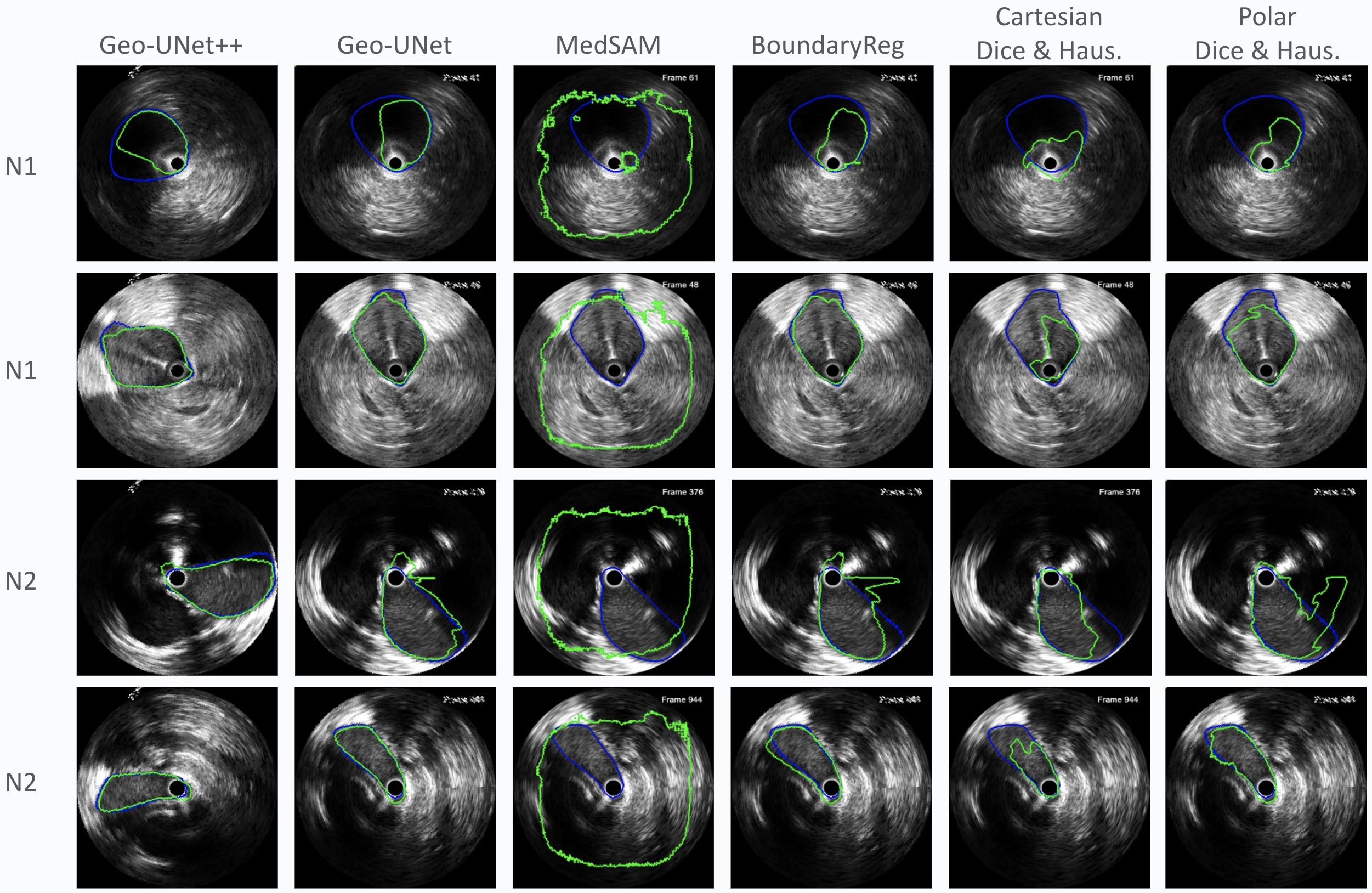}}
\caption{Additional segmentation result comparison across Geo-UNet++, Geo-UNet, and baselines for both N1 and N2 frames.} 
\label{examples}
\end{figure}

To quantify the generalization performance, we report the test-Dice and percentage of frames with major and minor diameter error within 0.25/0.5/0.75mm for N1 and N2 frames in the test subjects in Table~\ref{tab1} for all models. We observe that MedSAM~\cite{Ma_2024} severely under-performs all the conv-UNet frameworks trained on v-IVUS, due to an inability to meaningfully discern the lumen region without a more carefully curated manual prompt and generalization limitations. BoundaryReg underperforms Geo-UNet due to architectural differences and the lack of IVUS anatomy-rooted design decisions. The model trained in Cartesian space uniformly performs worse than all polar models, reinforcing our choice to use polar representations. The polar UNet trained on only pixel-wise segmentation performs worse than the Geo-UNet on several comparisons. Upon a qualitative examination (see Fig.~\ref{comparison}, ~\ref{examples} (supplementary)), the last two baselines can result in fragmented predictions with multiple components, as they are not constrained to predict a single lumen contour. This problem is not resolved by post-processing to choose the largest component given the heterogeneity across pullbacks and anatomical locations. Taking the output from the contour prediction branch inherently ensures a single prediction region. The combination of the two branches is effective as seen by comparing Geo-UNet and its ablated version without the pixel-wise prediction (Table~\ref{tab1}). Removing the re-weighting (CDFeLU) worsens performance on both N1 and N2 frames. Finally, Geo-UNet++, featuring continuity enhancement during inference, provides improvements in the estimates of the minor diameter, while maintaining the quality of the major diameter estimates for the N1 frames~\footnote{Errors on N2 major diams. remain above clinical precision despite slightly worsening}.  Overall, these observations make a strong case for adopting geometry-informed principles into the design of neural frameworks for lumen segmentation from v-IVUS imaging.

\medskip
\noindent\textbf{Future Directions:} Our framework can be easily extended beyond v-IVUS to applications with radial acquisitions/geometry such as multiple vessel boundary segmentation in arterial IVUS, retinal/airway OCT, laparoscopy, etc. Another natural extension of our framework is to 3D models that incorporate contextual information from adjacent frames~\cite{xu2022polar}. This is not entirely straightforward due to 1) Large shape variances among normal v-IVUS frames 2) Normal training frames constituting non-contiguous segments within a pullback between interspersed and anatomically distinct diseased frames and 3) Variable frame rates due to manual pullbacks. We envision that 3D rendering techniques such as Neural Implicit Functions~\cite{yariv2021volume} could potentially circumvent these issues.

\section{Conclusion}
We develop a novel geometry-informed neural model, Geo-UNet, for precise lumen segmentation on venous IVUS imaging for automated stent-sizing. The two-task design, i.e. lumen contour estimation and dense pixel prediction, ensures appropriate constraints per data geometry. The CDFeLU re-weighting allows us to unify the distinct prediction targets probabilistically and effectively mitigate spurious predictions. The inclusion of complementary losses provides sufficient regularization to ensure reliable and robust generalization across unseen pullbacks (patients) despite the modest dataset size. Finally, the inference time enhancement improves performance with negligible cost. Overall, Geo-UNet/Geo-UNet++ achieves a majority of clinical targets, with only a narrow gap in others, making it an attractive assistive tool for interventional specialists.

\bibliographystyle{splncs04}
\bibliography{lncs-bibliography}
\newpage

\section{Supplementary Material}

\end{document}